\documentclass[twocolumn,showpacs,preprintnumbers,amsmath,amssymb]{revtex4}
\usepackage[dvips]{graphicx}

\def\lsim{\buildrel {\textstyle <}\over {_\sim}}

\begin{document}
\title{Spin-chirality decoupling in the one-dimensional Heisenberg spin glass 
with long-range power-law interactions
}
\author{Dao Xuan Viet and Hikaru Kawamura}
\affiliation{Department of Earth and Space Science, Faculty of Science,
Osaka University, Toyonaka 560-0043,
Japan}
\date{\today}
\begin{abstract}
We study the issue of the spin-chirality decoupling/coupling in the ordering of the Heisenberg spin glass by performing large-scale Monte Carlo simulations on a one-dimensional Heisenberg spin-glass model with a long-range power-law interaction up to large system sizes. We find that the spin-chirality decoupling occurs for an intermediate range of the power-law exponent. Implications to the corresponding $d$-dimensional short-range model is discussed.
\end{abstract}
\maketitle

 The issue of the spin-glass (SG) ordering has been studied quite extensively for years, and continued to give an impact on surrounding areas \cite{review,Kawamura07,Kawamura09}. Nevertheless, the true nature of the ordering of SG magnets still remains elusive and controversial. Since the magnetic interaction in most of real SG materials is known to be nearly isotropic, they should be described as a first approximation by the isotropic Heisenberg model. Recently, consensus appears among various numerical works that the isotropic Heisenberg SG in three dimensions (3D) exhibits a finite-temperature transition, while the nature of the transition still remains controversial \cite{Kawamura92,HukuKawa00,HukuKawa05,VietKawamura09,Campos06,LeeYoung07,Fernandez09}. 

 It has been suggested in Ref.\cite{Kawamura92} that the model might exhibit an intriguing ``spin-chirality decoupling'' phenomenon, {\it i.e.\/}, the chirality exhibits the glass order at a temperature higher than the standard SG order, $T_{CG} > T_{SG}$ \cite{HukuKawa00,HukuKawa05,VietKawamura09}. Chirality is a multispin variable representing the handedness of the noncollinear or noncoplanar spin structures induced by frustration. By contrast, Refs.\cite{Campos06,LeeYoung07,Fernandez09} claim that the 3D Heisenberg SG does not exhibit such a spin-chirality decoupling, only a single transition which is simultaneously SG and chiral-glass (CG).

 To get deeper insight into the behavior in physical dimension $d=3$, it is often useful to study the phenomena by extending the dimensionality to general $d$ dimensions. In the limit of low $d$, the short-range (SR) Heisenberg SG exhibits only a $T=0$ transition in $d=1$. In $d=2$, recent calculations suggest that the vector SG model, either the three-component Heisenberg SG \cite{KawaYone03} or the two-component {\it XY\/} SG \cite{Weigel08}, exhibits a $T=0$ transition accompanied by the spin-chirality decoupling, {\it i.e.\/}, the CG correlation-length exponent $\nu_{CG}$ is greater than the SG correlation-length exponent $\nu_{SG}$. The spin-chirality decoupling associated with a finite-temperature transition could occur, if any, in $d\geq 3$. In the opposite limit of high $d$, the SR Heisenberg SG model in infinite dimensions $d\rightarrow \infty $ reduces to the mean-field (MF) model, {\it i.e.\/}, the Sherrington-Kirkpatrick (SK) model. The Heisenberg SK model is known to exhibit only a single finite-temperature SG transition, with no spin-chirality decoupling. In high but finite $d$, Monte Carlo (MC) result of Ref.\cite{ImaKawa03} suggested that the spin-chirality decoupling did not occur in $d=5$, but might occur in $d=4$. Reflecting an intrinsic difficulty in thermalizing large systems in high dimensions, however, the true situation still remains largely unclear.

 In the present paper, we attack the issue of the spin-chirality coupling/decoupling in the Heisenberg SG from a different perspective. Namely, we study a different type of Heisenberg SG model, {\it i.e.\/}, the one-dimensional (1D) Heisenberg SG with a long-range (LR) power-law interaction proportional to $1/r^{\sigma}$ ($r$ is the spin distance). In the limit of sufficiently large $\sigma \rightarrow \infty $, the model reduces to the standard $d=1$ model with a SR interaction. In the opposite limit of $\sigma \rightarrow 0$, the model reduces to an infinite-range model, {\it i.e.\/}, the SK model corresponding to $d=\infty $. Hence, varying $\sigma$ of the 1D LR model might be analogous to varying $d$ in the SR model. Indeed, this correspondence was supported by recent studies by Katzgraber and Young \cite{Katzgraber09} and  by Leuzzi {\it et al\/} \cite{Leuzzi08} for the Ising SG. These authors have suggested more detailed correspondence between  $d$ of the SR model and $\sigma$ of the 1D LR model, {\it e.g.\/}, (i) the upper critical dimension  $d=6$ corresponds to $\sigma=2/3$,  (ii) the lower critical dimension, which lies between $d=2$ and 3, corresponds to $\sigma=1$, and (iii)  $d=3$ corresponds to $\sigma \sim 0.9$.

 Advantages of studying such 1D models might be threefold. First, systems of large linear size $L$, never available in high dimensions, can be studied (up to $L=4096$ in the present calculation). Second, one can continuously change and even fine-tune the parameter $\sigma$ playing the role of effective ``dimensionality'', while it is impossible to continuously change the real dimensionality $d$ in the SR model. Hence, by studying the properties of the 1D model with varying $\sigma$, one might get  an overall picture concerning how the ``coupling vs. decoupling'' behavior depends on the effective dimensionality. Third, certain analytical results based on the renormalization-group (RG) calculations are available in 1D, which might be utilized in interpreting the numerical data. 

 Indeed, RG calculations, though did not take account of the possibility of the spin-chirality decoupling, suggested that the model exhibited a rich ordering behavior with varying $\sigma$ \cite{Kotliar82,Chang84}. For $\sigma \leq 2/3$, the Gaussian fixed point is stable and the model exhibited a finite-temperature SG transition of the MF type. For $2/3< \sigma < 1$,  a non-trivial LR fixed point becomes stable leading to a finite-temperature SG transition characterized by the non-MF exponents. In particular, the critical-point-decay exponent is determined solely by the power describing the spin-spin interaction, and is given by $\eta_{SG}=3- 2\sigma$ \cite{Kotliar82,Chang84}. For $\sigma \geq 1$, the SG transition occurs only at zero-temperature with $\eta_{SG}=1$.

\begin{figure}[ht]
\begin{center}
\includegraphics[scale=0.9]{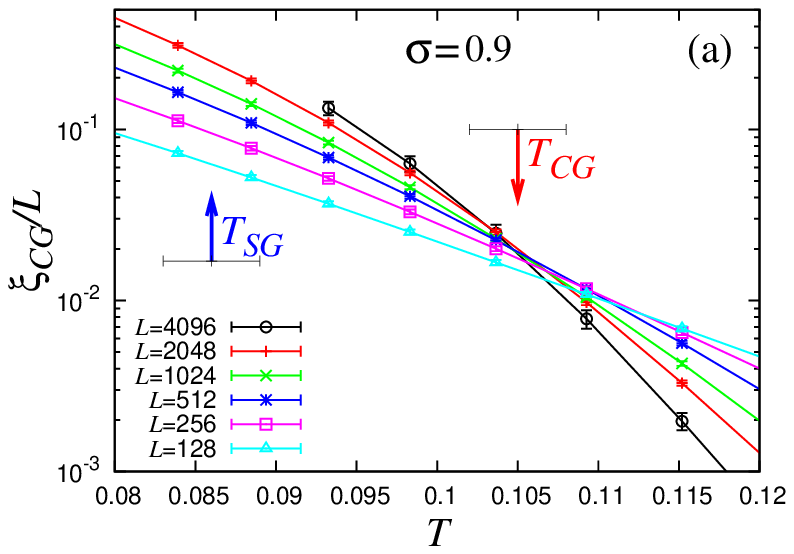}
\includegraphics[scale=0.9]{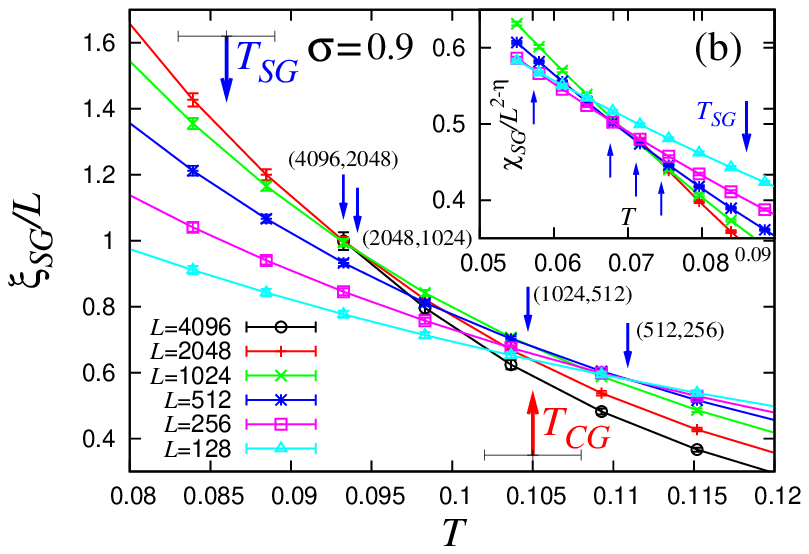}
\end{center}
\caption{
The correlation-length ratio versus the temperature for the chirality (a), and for the spin (b), for $\sigma=0.9$. The arrows indicate the bulk chiral-glass and spin-glass transition points. The inset of Fig.(b) represents the temperature dependence of the spin-glass susceptibility ratio, $\chi_{SG}/L^{2-\eta_{SG}}$ with $2-\eta_{SG}=2\sigma -1=2\times 0.9-1=0.8$.
}
\end{figure}
\begin{figure}[ht]
\begin{center}
\includegraphics[scale=0.9]{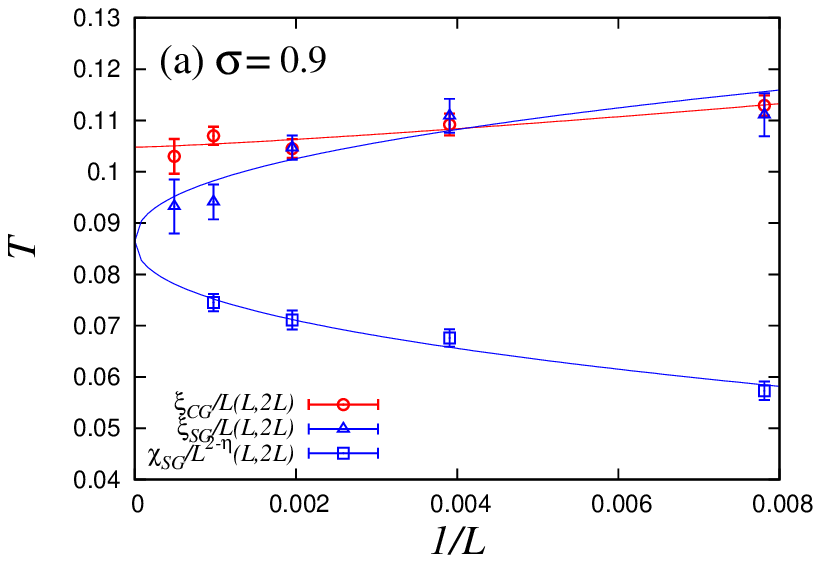}
\includegraphics[scale=0.9]{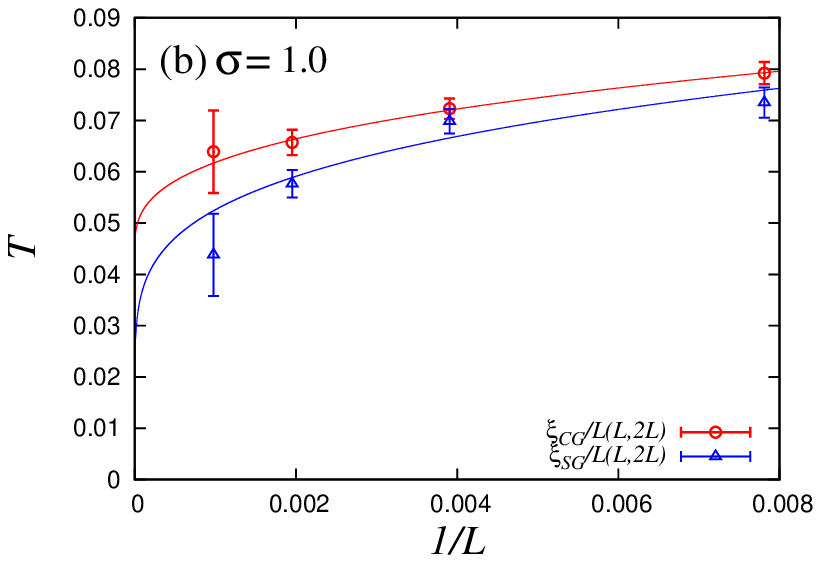}
\end{center}
\caption{
The (inverse) size dependence of the crossing temperatures of $\xi_{CG}/L$, $\xi_{SG}/L$, and $\chi_{SG}/L^{2-\eta_{SG}}$ for the case of $\sigma=0.9$ (a), and of $\sigma=1.0$ (b). Lines represent power-law fits with an exception of the SG data for $\sigma=1.0$ which is a logarithmic fit: See the text for details. 
}
\end{figure}

 Meanwhile, it remains to be seen how the spin-chirality decoupling arises in this 1D model with varying $\sigma$. Since the MF Heisenberg SK model does not show the spin-chirality decoupling, the spin-chirality decoupling associated with a finite-temperature transition should be realized, if any, only in the intermediate range of $\sigma$, near or below $\sigma=1$. Thus, we study here both the spin and the chiral orderings of the model by large-scale MC simulations, varying $\sigma$ in the range $0.7\leq \sigma \leq 1.1$, which spans the non-MF regime.  Our numerical results indicate that the model exhibits the spin-chirality decoupling in the range $0.8 \lsim \sigma \lsim 1.1$, while the usual spin-chirality coupling behavior occurs for $\sigma \lsim 0.8$.

 The Hamiltonian is the 1D classical Heisenberg model with a random LR power-law interaction $J_{ij}$,
\begin{equation}
{\cal H}=-\sum_{<ij>}J_{ij}\vec{S}_i\cdot \vec{S}_j\ \ ,
\label{eqn:hamil}
\end{equation}
where $\vec{S}_i=(S_i^x,S_i^y,S_i^z)$ is a three-component unit vector at the $i$-th site, and the $<ij>$ sum is taken over all spin pairs on the lattice once. The interaction $J_{ij}$ decays with a geometric distance $r_{ij}$ as a power-law, 
\begin{equation}
J_{ij}=C\frac {\epsilon_{ij}}{r_{ij}^{\sigma }},\ \ \ C=\surd {\frac {L}{\sum_{<ij>} r_{ij}^{-2\sigma}}},
\end{equation}
where $\epsilon_{ij}$ is an independent random Gaussian variable with zero mean and standard deviation unity. Periodic boundary condition is applied by placing $L$ spins on a ring. Then, the geometric distance between the spins at $i$ and $j$ is given by $r_{ij}=(L/{\pi })\sin(\pi \left| i-j \right|/L)$.

 We perform extensive MC simulations for various values of $\sigma$ in the range $0.7\leq \sigma \leq 1.1$ (a preliminary report for $\sigma=1.1$ was presented in Ref.\cite{MatsuKawa07}). We shall show below mainly the results for $\sigma=0.9$ and 1.0. The lattice sizes studied are $L$= 128, 256, 512, 1024, 2048, and also 4096 for some $\sigma$. Sample average is take over 896 (for $L\leq 2048$) and 256 ($L=4096$) independent bond realizations for $\sigma=0.9$, while 896 ($L\leq 1024$) and 256 ($L=2048$) for $\sigma=1.0$. We use a single-spin-flip heat-bath and an over-relaxation method combined with the temperature-exchange technique. The over-relaxation sweeps are repeated 5 times per every heat-bath sweep, which is set as our unit MC step. Equilibration is checked by monitoring: i) All the ``replicas'' travel back and forth many times (typically more than 10 times) along the temperature axis during the temperature-exchange process between maximum and minimum temperature points, whereas the relaxation due to single-spin flip is fast enough (both chiral and spin autocorrelation times about 20 MC steps or less) at the maximum temperature: (ii) All the measured quantities converge to stable values \cite{VietKawamura09}. 
%For further details, refer to Ref.\cite{VietKawamura09}.

 The local chirality at the $i$-th site $\chi_{i}$ is defined for three neighboring spins by the scalar $\chi_{i}=
\vec{S}_{i+1}\cdot (\vec{S}_i\times\vec{S}_{i-1})$. We simulate two independent copies of systems with identical interaction sets {$J_{ij}$} subject to mutually different random-number sequences and spin initial conditions, and measure $k$-dependent overlaps both for the spin and for the chirality. The $k$-dependent chiral overlap, $q_\chi(k)$, is defined as an overlap variable between the two replicas (1) and (2) by
\begin{equation}
q_\chi(k) =
\frac{1}{3N}\sum_{i=1}^N\sum_{\mu=x,y,z}
\chi_{i\mu}^{(1)}\chi_{i\mu}^{(2)}e^{i k  r_i}.
\end{equation}
From the chiral overlap, we calculate the CG susceptibility $\chi_{CG}$ via the second moment of its $k=0$ component, $\chi_{CG}=L[\langle | q_{\chi}(0)|^2 \rangle]$, where $\langle \cdots \rangle$ denotes a thermal average and $[\cdots ]$ an average over the bond disorder. Essentially the same definitions also apply to the spin except that an appropriate overlap becomes a tensor in spin space \cite{VietKawamura09}. Finite-size correlation length of the 1D LR model is then defined by
\begin{equation}
\xi = 
\frac{1}{2\sin(k_\mathrm{m}/2)}
\left( \frac{ [\langle q(0)^2 \rangle] }
{[\langle q(k_\mathrm{m})^2 \rangle] } -1 \right)^{1/(2\sigma -1)},
\end{equation}
for both SG and CG, where $k_\mathrm{m}= \frac{2\pi}{L}$ \cite{Katzgraber09,Leuzzi08}. 
% The reason for the power $1/(2\sigma -1)$ appearing in eq.(11) is that, at long wavelength, we expect a modified Ornstein-Zernike form for the LR model \cite{Mayor02,Katzgraber09,Leuzzi08}. 

 We begin with the case of $\sigma=0.9$ in the midst of the non-MF regime. In Fig.1, we show the correlation-length ratios for the chirality $\xi_{CG}/L$ (a), and for the spin $\xi_{SG}/L$ (b). These quantities are dimensionless so that the data of different $L$ should cross or merge at the expected transition point asymptotically for large $L$. For smaller sizes, the crossing temperatures $T_{cross}(L)$ of the spin $\xi_{SG}/L$ and of the chiral $\xi_{CG}/L$ are almost common, exhibiting only a weak $L$-dependence. For larger sizes, $T_{cross}(L)$ of the spin $\xi_{SG}/L$ shift down to lower temperatures, while those of the chiral $\xi_{CG}/L$ remain to be nearly $L$-independent, deviating from $T_{cross}(L)$ of the spin  $\xi_{SG}/L$. In Fig.2(a), we plot $T_{cross}(L)$ of the $\xi_{CG}/L$ curves between of the sizes $L$ and $2L$  as a function $1/L$, together with the corresponding ones of the $\xi_{SG}/L$ curves.

 While the SG and CG susceptibilities are dimensionfull quantities, they can be made dimensionless by dividing them by $L^{2-\eta}$ where $\eta$ is a critical-point-decay exponent. Generally, the exponent $\eta$ is not known in advance, but in the case of the present LR interaction, the SG exponent $\eta_{SG}$ is given by $\eta_{SG}=3-2\sigma$. Making using of this property, we plot in the inset of Fig.1(b) the temperature dependence of the SG susceptibility ratio $\chi_{SG}/L^{2-\eta_{SG}}$ where $2-\eta_{SG}=2\sigma-1=0.8$. As can be seen from the figure, the data of different $L$ exhibit a crossing behavior as expected for the dimensionless quantity.  For the CG susceptibility ratio, this type of analysis has only restricted utility because of the lack of our knowledge of the chiral $\eta_{CG}$ value.

\begin{figure}[ht]
\begin{center}
\includegraphics[scale=0.9]{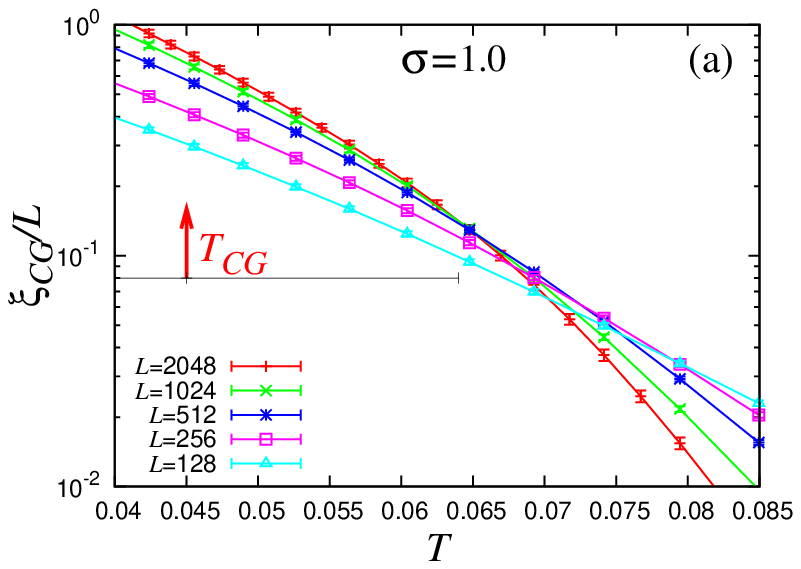}
\includegraphics[scale=0.9]{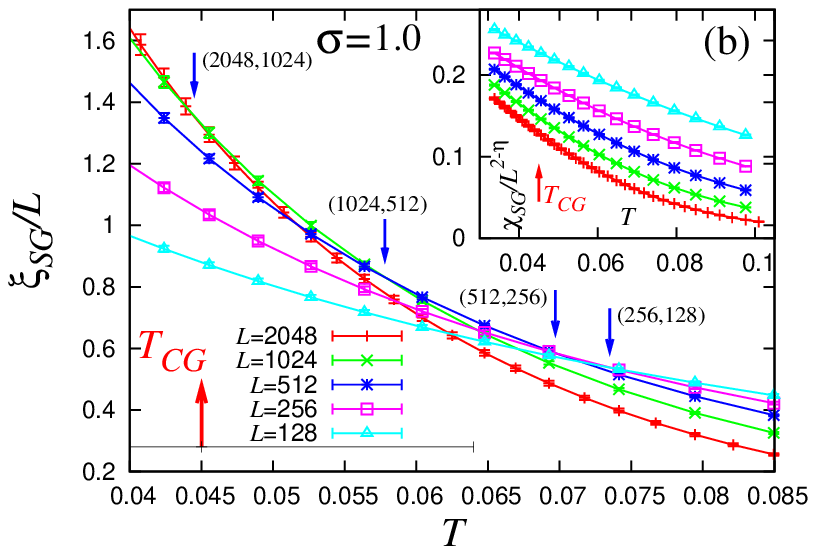}
\end{center}
\caption{
The correlation-length ratio versus the temperature for the chirality (a), and for the spin (b), for $\sigma=1.0$. The arrow indicates the bulk chiral-glass transition point. The inset of Fig.(b) represents the temperature dependence of the spin-glass susceptibility ratio, $\chi_{SG}/L^{2-\eta_{SG}}$ where $2-\eta_{SG}=2\sigma -1=2\times 1.0-1=1.0$.
}
\end{figure}
\begin{figure}[ht]
\begin{center}
\includegraphics[scale=0.9]{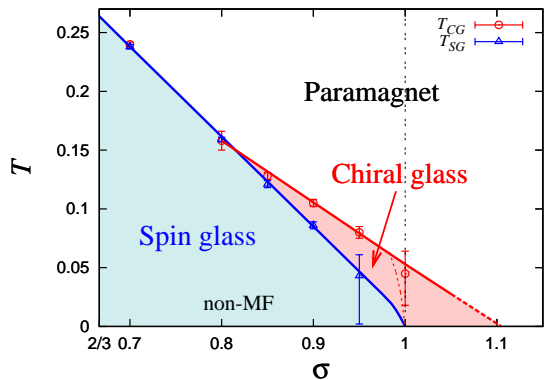}
\end{center}
\caption{
The $\sigma$ versus temperature phase diagram of the 1D Heisenberg spin glass with a LR power-law interaction decaying with a distance $r$ as $\propto r^{-\sigma}$. The red (blue) points are the chiral $T_{CG}$ (the spin $T_{SG}$) transition temperature. The spin-chirality decoupling occurs in the range $0.8 \lsim \sigma \lsim 1.1$, while more standard coupling behavior occurs in $\sigma \lsim 0.8$.
}
\end{figure}
 In order to estimate the bulk CG and SG transition temperatures quantitatively, we need to extrapolate $T_{cross}(L)$ to $L=\infty$. Such an extrapolation is done here on the basis of the relation, $T_{cross}(L)-T_{cross}(\infty)\approx c L^{-\theta}$ with $\theta=\nu^{-1}+\omega$, where $\nu$ and $\omega$ are the correlation-length and the leading correction-to-scaling exponents, respectively, while $c$ is a constant. For the SG, we perform a combined fit of $T_{cross}(L)$ of both $\xi_{SG}/L$ and of $\chi_{SG}/L^{2-\eta_{SG}}$, where a common $T_{cross}(\infty)$ and $\theta$ are assumed. The standard $\chi^2$-analysis then yields $T_{SG}=0.086\pm 0.003$ and $\theta =0.44 \pm 0.07$, with $\chi^2/$DOF$=1.24$ and the associated fitting probability $Q=0.29$. The smallness of the obtained error bar of $T_{SG}$ is due to the fact that the two independent $T_{cross}(L)$ are used in the fit, each approaching $T_{SG}$ either from above or below. For the CG, we have $T_{cross}(L)$ of $\xi_{CG}/L$ only, which yields $T_{CG}=0.105 \pm 0.003$ and $\theta =1.2 \pm 1.4$ ($\chi^2$/DOF=1.14 and $Q=0.32$). The smallness of the error bar of $T_{CG}$ is due to the fact that $T_{cross}(L)$ of $\xi_{CG}/L$ exhibits a nearly $L$-independent behavior. Hence, $T_{CG}$ turns out to be higher than $T_{SG}$ by about $20 \%$, suggesting that the spin-chirality decoupling occurs for $\sigma=0.9$.

On decreasing $\sigma$ from $\sigma=0.9$, one tends to approach the spin-chirality coupling regime. Indeed, for $\sigma=0.8$, we obtained via similar analyses (the data not shown here) $T_{CG} = 0.158\pm 0.008$ and $T_{SG} = 0.159\pm 0.002$, which suggests that the spin and the chirality might order simultaneously at $\sigma=0.8$. 

 By contrast, on increasing $\sigma$ from $\sigma=0.9$, ones approaches the $T_{SG}=0$ regime, with the upper-critical $\sigma$, $\sigma=1$. In Fig.3, we show the correlation-length ratios $\xi_{CG}/L$ (a) and $\xi_{SG}/L$ (b) for $\sigma=1.0$, together with the SG susceptibility ratio $\chi_{SG}/L^{2-\eta_{SG}}$.  In contrast the $\sigma=0.9$ case, $\chi_{SG}/L^{2-\eta_{SG}}$ does not show crossing in the investigated $T$ and $L$ range. In Fig.2(b), $T_{cross}(L)$ of $\xi_{CG}/L$ and of $\xi_{SG}/L$ as well as those of $\chi_{SG}/L^{2-\eta_{SG}}$ are plotted as a function of $1/L$. For the chirality, a fit of $T_{cross}(L)$ of $\xi_{CG}/L$ yields $T_{CG}=0.045^{+0.019}_{-0.027}$ and $\theta_{CG}=0.34\pm 0.34$ ($\chi^2$/DOF=0.15 and $Q=0.70$).  Hence, $T_{CG}$ is likely to be nonzero at $\sigma=1.0$. For the spin, the decreasing tendency of $T_{cross}(L)$ with $L$ becomes pronounced. In fact, a power-law fit becomes unstable here, leading to an indefinitely negative $T_{SG}$-value. Rather, a logarithmic fit of the form $T_{cross}(L) = b(\ln L + c)^{-\theta}$, expected for the $T=0$ transition at the upper-critical $\sigma$, yields an acceptable fit with $\theta \simeq 2.1$ ($\chi^2$/DOF$=3.88$ and $Q=0.0044$) as shown in Fig.2(b). This observation supports the  $T=0$ SG transition theoretically expected.

 Similar analyses are made for other values of $\sigma$ including $\sigma=0.7,0.8,0.85,0.95,1.1$. We then find the decoupling behavior with $T_{CG} > T_{SG}$ for $\sigma=0.85,0.95$, and the coupling behavior with $T_{CG} \simeq T_{SG}$ for $\sigma=0.7,0.8$ (details will be reported elsewhere). The obtained results are summarized in the $\sigma - T$ phase diagram in Fig.4. We tend to have larger error bars toward the upper-critical $\sigma=1$, because we have only one kind of $T_{cross}(L)$ and the data exhibit significant downward curvature there. The spin-chirality decoupling occurs in the range $0.8\lsim \sigma \lsim 1.1$. By contrast, the standard spin-chirality coupling behavior $T_{SG}=T_{CG}$ is realized for $\sigma \lsim 0.8$. The para-CG phase boundary might go beyond $\sigma=1$, touching the $T$-axis separately from the CG-SG phase boundary. 

 Although the correspondence between $\sigma$ of the 1D LR model and $d$ of the SR model is by no means exact, it might be interesting to deduce on the basis of Fig.4 the behavior of the $d$-dimensional Heisenberg SG with SR interactions. As mentioned, recent theoretical analyses suggest that the physical dimension $d=3$ corresponds to $\sigma$ slightly smaller than unity. Hence, our present conclusion that the spin-chirality decoupling occurs in the relevant range of $\sigma$, $0.8\lsim \sigma \lsim 1.1$, gives indirect support to the spin-chirality decoupling occurring in $d=3$ in the SR Heisenberg SG. Anyway, the phase diagram of Fig.4 serves to grasp an overall ordering behavior of the Heisenberg SG from a wider perspective.

 This study was supported by Grant-in-Aid for Scientific Research on Priority Areas ``Novel States of Matter Induced by Frustration'' (19052006). Numerical calculation was performed at the ISSP, Tokyo University, and at the YITP, Kyoto University. The authors are thankful to I.A. Campbell and H. Yoshino for useful discussion.

\end{document}